# Some fascinating series and their sums

Md Enamul Azim
University of Science and Technology, Chittagong
Md Mostofa Akbar and M Kaykobad
Department of Computer Science and Engineering
Bangladesh University of Engineering and Technology, Dhaka, Bangladesh
mostofa, kaykobad@cse.buet.ac.bd

**Abstract:** In this paper we present some interesting results involving summation of series in particular trigonometric ones. We failed to locate these results in existing literature or in the web like MathWorld (http://mathworld.wolfram.com/) nor could we derive them using software for analytical computation like *Maple*. The identities are beautiful and involve finite series.

## 1. Introduction

Mathematical identities are quite interesting in their own right. Quite often they are not only beautiful, they often go beyond imagination in their forms and simplicity with which they appear. Most of the trigonometric functions are evaluated by summing an appropriate series. For example, the value of $\pi$ is computed using $x = \frac{\pi}{4}$ in Leibnitz equation $\tan^{-1} x = x - \frac{1}{3x^3} + \frac{1}{5x^5} - \frac{1}{7x^7} + ...$ . Other trigonometric functions are also computed using appropriate series. By far the most frequently accessed trigonometric series appears in Fourier series. Knuth's series sum is expressible in terms of Riemann zeta function in the following way

$$\sum_{k=1}^{\infty}\left(\frac{k^k}{k!e^k} - \frac{1}{\sqrt{2\pi k}}\right) = -\frac{2}{3} - \frac{1}{\sqrt{2\pi}}\zeta(\frac{1}{2})$$

Ramanujan discovered many beautiful expressions as sums of many series. Weisstein[2] has created a he repository of important series. In this paper we present some series and deduce their sums. These deductions are simple and straight forward although could not be found in existing literature and in the web.

## 2. Main Results

In this section we present some new identities including trigonometric functions that do not appear to be available in the existing literature. At the same time these identities appear to be too beautiful to remain unwarranted for publication. We present the identities in the form of theorems as below.

**Theorem 2.1.** $\sum_{k=0}^{2k \leq n} (-1)^k \binom{n}{2k} x^{n-2k} y^{2k} = (x^2 + y^2)^{\frac{n}{2}} \cos(n \tan^{-1} \frac{y}{x})$

$\sum_{k=0}^{2k < n} (-1)^k \binom{n}{2k+1} x^{n-(2k+1)} y^{2k+1} = (x^2 + y^2)^{\frac{n}{2}} \sin(n \tan^{-1} \frac{y}{x})$

**Proof:**

$(x+iy)^n = \sum_{k=0}^{n} \binom{n}{k} x^{n-k} (iy)^k$

$= \sum_{k=0}^{2k \leq n} \binom{n}{2k} (-1)^k x^{n-2k} y^{2k} + i \sum_{k=0}^{2k \leq n} \binom{n}{2k} (-1)^k x^{n-2k} y^{2k}$ (2.1)

But,

$(x+iy)^n = (\sqrt{x^2 + y^2} e^{i \tan^{-1} \frac{y}{x}})^n = (x^2 + y^2)^{\frac{n}{2}} e^{in \tan^{-1} \frac{y}{x}}$ (2.2)

So,

$(x^2 + y^2)^{\frac{n}{2}} \cos(n \tan^{-1} \frac{y}{x}) = \sum_{k=0}^{2k \leq n} (-1)^k \binom{n}{2k} x^{n-2k} y^{2k}$ (2.3)

and,

$(x^2 + y^2)^{\frac{n}{2}} \sin(n \tan^{-1} \frac{y}{x}) = \sum_{k=0}^{2k < n} (-1)^k \binom{n}{2k+1} x^{n-(2k+1)} y^{2k+1}$ (2.4)

Squaring (2.3) and (2.4) and gives us

$(\sum_{k=0}^{2k \leq n} (-1)^k \binom{n}{2k} x^{n-2k} y^{2k})^2 + (\sum_{k=0}^{2k < n} (-1)^k \binom{n}{2k+1} x^{n-(2k+1)} y^{2k+1})^2$

$= (x^2 + y^2)^n$ (2.5)

**Corollary 2.1** $2^{\frac{n}{2}} \cos(\frac{n\pi}{4}) = \sum_{k=0}^{2k \leq n} (-1)^k \binom{n}{2k}$ (2.6)

and

$2^{\frac{n}{2}} \sin(\frac{n\pi}{4}) = \sum_{k=0}^{2k < n} (-1)^k \binom{n}{2k+1}$ (2.7)

We get (2.6) and (2.7) by putting $x = y$ respectively in (2.3) and (2.4)
Now from (2.6) and (127) we get

$$\sqrt{2} = \sqrt{\frac{\sum_{k=0}^{2k \leq n}(-1)^k \binom{n}{2k}}{\cos(\frac{n\pi}{4})}} = \sqrt{\frac{\sum_{k=0}^{2k<n}(-1)^k \binom{n}{2k+1}}{\sin(\frac{n\pi}{4})}} \qquad (2.8)$$

From (2.8) we have

$$\tan(\frac{n\pi}{4}) = \frac{\sum_{k=0}^{2k<n}(-1)^k \binom{n}{2k+1}}{\sum_{k=0}^{2k \leq n}(-1)^k \binom{n}{2k}} \qquad (2.9)$$

(2.9) implies that $\tan(\frac{n\pi}{4})$ is a rational number for integer n. It can be checked that for n=3 numerator is 2 whereas denominator is -2 giving us the correct values of -1.

**Theorem 2.2** Let $x_k, y_k$ be integers for $k = 1, 2, ..., n$. Then both

$$\prod_{k=1}^{n}\sqrt{x_k^2 + y_k^2} \cos(\sum_{k=1}^{n}\tan^{-1}\frac{y_k}{x_k}) \text{ and } \prod_{k=1}^{n}\sqrt{x_k^2 + y_k^2} \sin(\sum_{k=1}^{n}\tan^{-1}\frac{y_k}{x_k}) \text{ are integers.}$$

**Proof:** Let $z_k = x_k + iy_k$ then

$$\prod_{k=1}^{n} z_k = \prod_{k=1}^{n}(x_k + iy_k) = \prod_{k=1}^{n}\sqrt{x_k^2 + y_k^2} e^{i\tan^{-1}\frac{y_k}{x_k}} = \prod_{k=1}^{n}\sqrt{x_k^2 + y_k^2} \times \prod_{k=1}^{n} e^{i\tan^{-1}\frac{y_k}{x_k}} =$$

$$= \prod_{k=1}^{n}\sqrt{x_k^2 + y_k^2} \times e^{i\sum_{k=1}^{n}\tan^{-1}\frac{y_k}{x_k}} \qquad (2.10)$$

From (2.10) we can say that both real and imaginary parts of the product are integers.

Hence $\prod_{k=1}^{n}\sqrt{x_k^2 + y_k^2} \cos(\sum_{k=1}^{n}\tan^{-1}\frac{y_k}{x_k})$ and $\prod_{k=1}^{n}\sqrt{x_k^2 + y_k^2} \sin(\sum_{k=1}^{n}\tan^{-1}\frac{y_k}{x_k})$ are both integers.

**Theorem 1.3** $(\sum_{k=1}^{n}\sin kx)^2 + (\sum_{k=1}^{n}\cos kx)^2 = \frac{1-\cos nx}{1-\cos x}$

In order to prove the theorem we have the following lemma.

**Lemma 2.1** $1 + 2\sum_{k=1}^{m}\cos kx = \frac{\cos mx - \cos(m+1)x}{1-\cos x}$

**Proof:** We prove the lemma by induction. For m=1

$$\text{RHS} = \frac{\cos x - \cos 2x}{1-\cos x} = \frac{\cos x - (2\cos^2 x - 1)}{1-\cos x} = 1 + 2\cos x = \text{LHS}$$

Let us assume it to be true for $< m$. Now for $m$ we have

$$\text{LHS} = 1 + 2\sum_{k=1}^{m} \cos kx = 1 + 2\sum_{k=1}^{m-1} \cos kx + 2\cos mx$$

$$= \frac{1-\cos(m-1)x}{1-\cos x} + 2\cos mx$$

$$= \frac{\cos(m-1)x - \cos mx + 2\cos mx - 2\cos x \cos mx}{1-\cos x}$$

$$= \frac{\cos mx - \cos(m+1)x}{1-\cos x} = \text{RHS}$$

Now we come back to the proof of Theorem 2.3

**Proof:** Again we prove this theorem by induction.

For $n = 1$, LHS$=\sin^2 x + \cos^2 x = 1 = \dfrac{1-\cos x}{1-\cos x} = $ RHS

Let us assume that the theorem holds true for $< n$. Now to prove for $n$ we have

$$\text{LHS} = (\sum_{k=1}^{n} \sin kx)^2 + (\sum_{k=1}^{n} \cos kx)^2 = (\sum_{k=1}^{n-1} \sin kx + \sin nx)^2$$

$$+ (\sum_{k=1}^{n-1} \cos kx + \cos nx)^2$$

$$= (\sum_{k=1}^{n-1} \sin kx)^2 + (\sum_{k=1}^{n-1} \cos kx)^2 + 2\sin nx \sum_{k=1}^{n-1} \sin kx$$

$$+ 2\cos mx \sum_{k=1}^{n-1} \cos kx + 1$$

$$= \frac{1-\cos(n-1)x}{1-\cos x} + \sum_{k=1}^{m-1}\{\cos(n-k)x - \cos(n+k)x\}$$

$$+ \sum_{k=1}^{m-1}\{\cos(n-k)x - \cos(n+k)x\} + 1$$

$$= \frac{1-\cos(n-1)x}{1-\cos x} + 2\sum_{k=1}^{n-1} \cos(n-k)x + 1$$

$$= \frac{1-\cos(n-1)x}{1-\cos x} + 2\sum_{k=1}^{n-1} \cos kx + 1$$

$$= \frac{1-\cos(n-1)x}{1-\cos x} + \frac{\cos(n-1)x - \cos nx}{1-\cos x} = \frac{1-\cos nx}{1-\cos x}$$

So this proves the theorem.

**Theorem 2.4** $\sum_{k=1}^{n} \frac{1}{2^k} \tan \frac{x}{2^k} = \frac{1}{2^n} \cot \frac{x}{2^n} - \cot x$

**Proof:** Again we use induction to prove this identity.

For $n = 1$, RHS $= \frac{1}{2} \cot \frac{x}{2} - \cot x = \frac{1}{2} \cot \frac{x}{2} - \frac{\cot^2 \frac{x}{2} - 1}{1 + \cot \frac{x}{2}} = \frac{1}{2} \tan \frac{x}{2} =$ RHS

Let it be true for $< n$. Now for n

$$\text{LHS} = \sum_{k=1}^{n} \frac{1}{2^k} \tan \frac{x}{2^k} = \sum_{k=1}^{n-1} \frac{1}{2^k} \tan \frac{x}{2^k} + \frac{1}{2^n} \tan \frac{x}{2^n}$$

$$= \frac{1}{2^{n-1}} \cot \frac{x}{2^{n-1}} - \cot x + \frac{1}{2^n} \tan \frac{x}{2^n}$$

$$= \frac{1}{2^{n-1}} \left( \frac{\cot^2 \frac{x}{2^n} - 1}{2 \cot \frac{x}{2^n}} \right) - \cot x + \frac{1}{2^n} \tan \frac{x}{2^n}$$

$$= \frac{1}{2^n} \cot \frac{x}{2^n} - \frac{1}{2^n} \tan \frac{x}{2^n} - \cot x + \frac{1}{2^n} \tan \frac{x}{2^n}$$

$$= \frac{1}{2^n} \cot \frac{x}{2^n} - \cot x$$

This proves the theorem.

We have the following interesting identity.

**Theorem 2.5.**

$$\sum_{k=1}^{n} \tan kx \tan(k+1)x = -(n+1) + \frac{\tan(n+1)x}{\tan x} \qquad (2.11)$$

**Proof:** Let

$$T_n(x) = \sum_{k=1}^{n} \tan kx \text{ then } T_n'(x) = \sum_{k=1}^{n} k \sec^2 kx = \sum_{k=1}^{n} k \tan^2 kx + \frac{n(n+1)}{2}$$

$$T_n'(-x) = \sum_{k=1}^{n} k \tan^2 kx + \frac{n(n+1)}{2}, \, T_n(-x) = -\sum_{k=1}^{n} \tan kx$$

$$\tan kx = \tan(k+1-1)x = \frac{\tan(k+1)x - \tan x}{1 + \tan(k+1)x \tan x}$$

$$\tan kx + \tan kx \tan x \tan(k+1)x = \tan(k+1)x - \tan x$$

Now adding up over $k$ we have

$$\sum_{k=1}^{n}\tan kx+\sum_{k=1}^{n}\tan kx\tan x\tan(k+1)x=\sum_{k=1}^{n}\tan(k+1)x-\sum_{k=1}^{n}\tan kx$$

$$\Rightarrow T_n(x)+\sum_{k=1}^{n}\tan kx\tan x\tan(k+1)x=T_n(x)-T_1(x)+\tan(n+1)x-n\tan x$$

$$\Rightarrow \sum_{k=1}^{n}\tan kx\tan x\tan(k+1)x=-T_1(x)+\tan(n+1)x-n\tan x$$

$$=-\tan x+\tan(n+1)x-n\tan x$$

$$\Rightarrow \sum_{k=1}^{n}\tan kx\tan(k+1)x=-(n+1)+\frac{\tan(n+1)x}{\tan x}$$

The result can be easily verified for the case $n=2$, whence

$$\tan x\tan 2x=\frac{\sin x\sin 2x}{\cos x\cos 2x}=\frac{2\tan^2 x}{1-\tan^2 x}=-2+\frac{2}{1-\tan^2 x}$$

$$=-2+\frac{2\tan 2x}{\tan 2x(1-\tan^2 x)}$$

$$=-2+\frac{2\dfrac{2\tan x}{1-\tan^2 x}}{\dfrac{2\tan x}{1-\tan^2 x}(1-\tan^2 x)}=-2+\frac{\tan 2x}{\tan x}=-(1+1)+\frac{\tan 2x}{\tan x}$$

In the same way the following identity can also be established.

**Theorem 2.5**

$$\sum_{k=1}^{n}\tan kx\left[\tan(k-1)x+\tan(k+1)x\right]=\frac{1}{\tan x}\left[\tan nx+\tan(n+1)x\right]-1-2n$$

### 3. Conclusion

The identities presented in section 2 involve sum of finite trigonometric series that deserve attention simply since these are beautiful.

### Acknowledgement

These and many more interesting and beautiful formulae were derived by the first author under the loose supervision of the other coauthors. Very unfortunately this young talented computer graduate from Bangladesh University of Engineering and Technology succumbed to a tragic death at his early twenties.

**References**

1. Knuth, D. E. "Problem 10832." *Amer. Math. Monthly* **107**, 863, 2000.
2. Weisstein, Eric W.  *MathWorld*--A Wolfram Web Resource. http://mathworld.wolfram.com/FourierSeries.html